# Semantic Non-Assembly: Privacy by Architectural Inertness Under Component Exposure

**Sam Ryan**
Novel Systems Engineering LLC
Portland, OR, United States
sam@novelsystems.me
ORCID: 0009-0000-1862-0417

**ABSTRACT**

Existing privacy frameworks emphasize confidentiality, access control, appropriate information flow, or statistical disclosure limitation. We introduce a complementary class of privacy guarantee (Semantic Non-Assembly) in which privacy is characterized not by the difficulty of achieving exposure but by the information yield of exposure when it occurs. SNA prevents evaluation of a designated predicate by preventing any sub-threshold coalition from assembling a sufficient assignment to its input domain. An architecture satisfies Semantic Non-Assembly when no coalition of fewer than a defined threshold of components can assemble such an assignment: complete exposure and decryption of any sub-threshold component yields no actionable data. In the base protocol, the guarantee is structural: it operates through architecture, not policy, and its privacy properties degrade predictably under component compromise rather than collapsing at a single point. The reference instantiation combines this structural guarantee with audited organizational constraints, as characterized in Appendix A. This paper formalizes the guarantee and establishes four ProVerif-verified properties: Device Non-Correlation, Registry Observer Non-Identification, Submission Server Blindness, and Active Defense Gate correctness, the first three through a two-channel provenance architecture. The Birthmark Standard instantiates the guarantee on constrained capture hardware, demonstrating deployability where ZK-based approaches are computationally infeasible. All formal properties and scope limitations are documented in Appendix A.

## 1 INTRODUCTION

Component compromise is no longer an exceptional event. Credential theft, insider access, and infrastructure breach occur at sufficient frequency that designing systems around their prevention is insufficient. A system whose privacy guarantee degrades to zero on single-component exposure has, in practice, no privacy guarantee at all. The prevailing frameworks (confidentiality, access control, information flow control, statistical disclosure limitation) address how to restrict access to sensitive information that exists somewhere in the system. None addresses the case where the armor has been pierced. All assume a store to protect, and all fail when that store is exposed. What is needed is a class of system that concedes nothing on component exposure, because there is nothing to concede.

A provenance architecture illustrates the requirement: a device must be linkable to authenticated content, but only under specific authorized conditions. The privacy challenge is not to hide the linkage but to ensure no component accumulates sufficient information to perform it without authorization. This paper asks: can a system be architected such that complete, authorized exposure of any single component (including full decryption of its contents) yields no information sufficient to execute the target identification function? In a threat environment where component exposure is probable rather than merely possible, the yield of exposure is a more actionable security property than its difficulty. Semantic Non-Assembly is the class of guarantee that answers this question affirmatively.

The contributions of this paper are:

(1) Semantic Non-Assembly as a formalized class of privacy guarantee: a structural property characterizing privacy by the information yield of component exposure, not the difficulty of achieving it, distinct from confidentiality, access control, information flow control, and statistical disclosure limitation (Section 1.1, Section 4).

(2) A two-channel provenance architecture instantiating the guarantee, with executable ProVerif proofs establishing Device Non-Correlation (Property A), Registry Observer Non-Identification (Property B), and Submission Server Blindness (Property C). Properties A, B, and C are mechanized in ProVerif; the reference instantiation uses ECIES (asymmetric IND-CCA2) for credential encryption, and the ProVerif model directly reflects this deployment. Active defense gate properties (Property D: gate correspondence, purge-state secrecy, chain dependency, chain integrity, and opaque index secrecy) are formally verified in BirthmarkD.pv and documented in Appendix A.4 (all properties scoped in Appendix A). Table 3 (Section 7.2) documents what each component compromise yields and does not yield; the guarantee degrades predictably under compromise rather than collapsing at a single point.

(3) A formal characterization of the privacy cost of adding forensic capability: the cost is bounded, predictable, and documented in Section 5.

(4) The Birthmark Standard as a reference instantiation demonstrating deployability on constrained capture hardware (Section 6). Full implementation details are documented in the extended technical report [4].

### 1.1 Semantic Non-Assembly

SNA prevents evaluation of a designated predicate by preventing any sub-threshold coalition from assembling a sufficient assignment to its input domain. The term "semantic" here refers to the meaning level of information (what the data enables with respect to f) rather than to semantic security in the cryptographic sense. Formally: an architecture satisfies semantic non-assembly with respect to predicate f and threshold t if no coalition of fewer than t+1 components can assemble a state from which f is computable. A sufficient state is any combination of information fields whose values together enable execution of f; semantic non-assembly requires that no sub-threshold coalition can assemble such a combination. For device attribution in media provenance, f is the function that links a specific capture device to specific authenticated content; a sufficient state is any combination of device identity and content record that enables execution of that linkage. The threshold t formalizes the adversary assumption: an architecture satisfying semantic non-assembly at t=1 remains private under any single-component compromise; larger t tolerates broader coordinated compromise. Equivalently: complete exposure and full decryption of any component below threshold t yields no actionable data with respect to f, because the sufficient state for f cannot be assembled from that component's contents alone.

The mechanism is architectural rather than procedural. A policy prohibition on co-locating sufficient information is not Semantic Non-Assembly; it is an access control rule applied to a system that has the capability to violate its privacy obligations. Semantic Non-Assembly requires that the architecture make co-location structurally impossible below the compromise threshold by design, not just by trust. This distinction matters for the adversary model: policy-based separation fails under insider threat or operator compromise; structural separation fails only under coordinated multi-component compromise at or above threshold t.

Because SNA is defined at the level of predicate computability over a distributed information schema (rather than access restriction, flow control, or inference bounds) it is distinct from each major existing privacy mechanism:

**Confidentiality** restricts who can access information that exists in one place. Semantic Non-Assembly prevents the sufficient information set from existing in one place. A system with strong confidentiality can still be vulnerable to single-component compromise if that component holds sufficient information; Semantic Non-Assembly removes the sufficient-information set from the single-component threat surface entirely.

**Access control** mediates operations on existing stores. Semantic Non-Assembly eliminates the store. Access control failure (misconfigured permissions, insider threat, key compromise) yields the information regardless. Semantic Non-Assembly failure requires simultaneous failure of t+1 components.

**Information flow control** restricts propagation paths for data that exists at a source, using runtime policy enforcement or label-based tracking. SNA operates at design time: no component ever holds the sufficient information set, so there is no flow to restrict or label. An IFC policy can be bypassed under insider threat or misconfiguration; structural SNA cannot be bypassed below threshold t because the target state is architecturally absent, not just prohibited.

**Statistical disclosure limitation** bounds what can be inferred from released data by adding noise, suppression, or generalization. The object of protection is a dataset that exists somewhere. SNA prevents the sufficient information set from existing at any sub-threshold component: an adversary with single-component access has no dataset to apply inference to, not just bounded inference from one.

Data minimization mitigates the consequences of exposure by reducing what exists to be taken. SNA prevents the exposure from being actionable regardless of what exists: it is a defense layer that operates before damage needs to be calculated, not a tighter specification of minimization policy.

The class of guarantee applies to systems where (a) the target identification function is defined and bounded, (b) the system designer controls which components hold which information fields, and (c) the privacy requirement is that the function be performable only under authorized conditions, not that it be permanently prevented. Provenance architectures are the canonical case: device-to-content linkage is the system's purpose, and privacy requires that linkage be authorizable but not passively available.

### 1.2 Minimal Instantiation: Federated Identity Verification

A federated identity system must solve a constrained verification problem: confirm that a user holds a valid credential, without revealing to the service which user it is, and without revealing to the identity provider which service the user is accessing. Both directions of the correlation must be blocked simultaneously. Blocking only one yields a system where either the identity provider accumulates a map of user activity across services, or the service accumulates a map of identities behind sessions. Neither partial solution satisfies the privacy requirement.

Four parties: an identity provider (Party A, holding user credentials), a service provider (Party B, holding session records), a routing authority (Party C, holding the mapping from opaque destination tokens to service endpoints), and a verification exchange. When a user requests access, they encrypt their credential under Party A's key and send the ciphertext to the exchange, along with an opaque destination token pre-registered with Party C. The exchange routes the encrypted credential to Party A, which decrypts it, verifies membership, and returns a signed authorization. Party A receives no information about the destination. The exchange passes the opaque token to Party C, which resolves the endpoint and forwards the session token to Party B. Party B sees an authorized token with no recoverable user identity. The exchange sees an encrypted credential it cannot decrypt and an opaque token it cannot resolve. No party ever holds both a plaintext user identity and a service access record simultaneously.

After the session completes, f(user, service) = "did this user access this service?" is not computable from any single party's state. Party A holds credential records with no associated service destinations. Party B holds session records with no associated user identities. Party C holds destination mappings with no associated credentials or session records. The exchange holds neither surface in decryptable form. Full exposure of any component yields no actionable data with respect to f. Component contents are not obscured; they are simply insufficient.

Access control explains why Party A should not log which services users access. Information flow control explains why user identities should not propagate to Party B. Neither framework can state the zero-yield property. Both presuppose a store to protect or a flow to restrict. The sufficient state here (the pairing of a plaintext user identity with a service access record) never exists at any component. SNA describes the class of systems structured so that no component ever holds it.

The threshold parameter follows directly. At $t=1$ the base architecture holds: no single-component compromise reveals the user-service pairing. Collusion between any two parties is the $t=2$ failure mode, requiring coordinated access across organizational boundaries. A back-lookup extension enabling authorized disclosure under legal process follows the same four-party structure as Section 5, with the privacy cost bounded identically.

### 1.3 Scope

The architecture verifies that content originated from a registered capture device. It does not verify scene truthfulness, content accuracy, staging authenticity, or editorial decisions. Privacy guarantees are computational rather than information-theoretic. Section 4.1 defines the adversary model precisely, and Section 7.1 characterizes the tradeoff relative to ZK-proof approaches. This paper establishes the SNA definition, the channel separation proofs (Properties A, B, C), and the architecture independently. Active defense gate proofs (Property D) are documented in Appendix A.4. Hardware benchmarks and the full governance specification are documented in the extended technical report [4], which is an arXiv preprint not independently peer-reviewed at time of this submission.

## 2 RELATED WORK

Privacy engineering frameworks provide the landscape against which Semantic Non-Assembly is positioned. Confidentiality and access control are foundational [7]. Contextual integrity (Nissenbaum [14]) frames privacy as appropriate information flow relative to context norms. SNA is distinct: contextual integrity asks whether a given flow is appropriate relative to the norms of its context; SNA asks whether the sufficient state for a target function can be assembled at all, independent of whether any individual flow is norm-violating. A system can satisfy contextual integrity in every flow and still fail SNA if a single component accumulates a sufficient state through individually appropriate flows. Differential privacy (Dwork et al. [15]) provides formal guarantees for statistical data releases. k-anonymity and its successors (l-diversity, t-closeness) address quasi-identifier-based re-identification in tabular data. Purpose limitation under GDPR and similar frameworks imposes policy-level constraints on information use. None of these frameworks addresses the structural prevention of sufficient-information assembly in systems where the identification function is authorized but must not be passively available.

**Embedded Provenance.** C2PA embeds cryptographically signed manifests in media files [1], but social media platforms routinely strip metadata through compression and format conversion, severing provenance at distribution. Storing verification records externally makes survival of in-band metadata loss a structural property rather than a platform dependency. The failure modes differ: C2PA's provenance integrity depends on metadata survival through the distribution chain; this architecture's privacy properties depend on audited constraint preservation across independently operated infrastructure components.

**Zero-Knowledge Proofs for Authentication.** ZK-SNARKs and ZK-STARKs enable proof of credential possession without credential disclosure [8]. Applied to device authentication, they allow a capture device to prove membership in a set of registered devices without identifying which member it is. The formal privacy guarantees are strong: correlation is prevented even against an adversary who compromises the credential validator. The deployment barrier is computational: ZK proof generation on ARM-class embedded hardware requires seconds to minutes per proof at realistic constraint counts [5], incompatible with capture-path latency requirements. This architecture targets the deployment contexts where this barrier makes ZK-proof approaches impractical. The failure modes differ: ZK systems rely primarily on computational hardness assumptions; this architecture relies on continuously audited structural constraints combined with standard cryptographic assumptions. Both are reasonable under their respective deployment conditions.

**Consortium Blockchains.** Substrate-based permissioned blockchains provide controlled validator sets without cryptocurrency exposure [3]. We use a consortium blockchain as an append-only registry for content hashes, providing tamper-resistance and public verifiability without token economics.

**Split-Knowledge Systems and Privacy Engineering.** The principle that no single party should hold sufficient information to perform a sensitive operation is established in information security through dual-control and split-knowledge architectures [7], used in financial institutions, key management systems, and cryptographic escrow. A split-knowledge system partitions an element $x \in X$ across shares. An SNA system partitions X itself across components, such that no sub-threshold coalition holds a subset of fields sufficient to assemble a state from which f is computable. More precisely: split-knowledge architectures partition a single algebraic object (a key or secret value) into shares distributed across parties, such that reconstruction requires a threshold of shares. The object of protection is that algebraic element; the mechanism is threshold reconstruction. Semantic Non-Assembly operates on a typed information schema rather than a single algebraic object: the object of protection is a predicate's input domain (the set of information fields whose combination constitutes a sufficient input to the identification function), and the mechanism is preventing that domain from being assembled at any component below threshold t. A threshold scheme over a device credential and a content hash would combine them into a shared secret requiring t-of-n parties to reconstruct; Semantic Non-Assembly ensures the two fields are never co-located, making the threshold reconstruction question moot. The distinction matters under insider threat: a t-of-n threshold scheme fails when t colluding parties reconstruct the secret; Semantic Non-Assembly fails only when t+1 components are simultaneously compromised, and even then the adversary must assemble the sufficient set across components rather than reconstruct a pre-existing value.

Conventional data minimization seeks to reduce the amount of retained information. Semantic Non-Assembly instead minimizes the information available to any architectural component with respect to a designated identification function. The objective is structural absence of any sufficient state from which the function is computable, independent of record size. In this sense, Semantic Non-Assembly operationalizes data minimization at the level of predicate computability rather than record size. The GDPR's data minimization principle (Article 5(1)(c)) requires that personal data be adequate, relevant, and limited to what is necessary; SNA operationalizes this as a structural property rather than a policy constraint, ensuring that no component accumulates data beyond what its function requires. Applying these established principles to public media provenance is, to our knowledge, not previously described in the literature.

The resulting system is openly queryable while remaining semantically inert to observers who do not already possess the underlying media, all within the computational budget of a camera image signal processor (ISP).

**Multi-Party Computation and t-Privacy.** The zero-yield-at-threshold property recalls t-privacy in multi-party computation [13]: in a t-private MPC protocol, no coalition of fewer than t+1 parties can learn anything beyond what their inputs and outputs reveal. The distinction from Semantic Non-Assembly is structural. MPC t-privacy operates on computational protocols that distribute shares of input values across parties, such that no sub-threshold coalition can reconstruct the function's inputs from its shares; the object of protection is the computational secrecy of input values during protocol execution. Semantic Non-Assembly operates on information architecture: the object of protection is the assembly of a sufficient state across components, and the guarantee holds independent of any protocol executed over that architecture. A system can satisfy SNA without executing any MPC protocol; an MPC protocol does not automatically confer SNA on the infrastructure it runs on. The two properties are orthogonal rather than hierarchical.

# 3 ARCHITECTURE

## 3.1 System Overview

The architecture involves four parties: a capture device that generates a content hash and a device credential; a credential validator (an authority that can confirm whether a credential corresponds to a registered device); a submission server that receives device submissions and routes credentials to the validator; and a verification registry (a consortium blockchain) that stores content hashes for public query.

Device registration is a prerequisite outside the scope of this architecture; credential provisioning is deployment-specific and may involve manufacturer enrollment, device initialization protocols, or similar mechanisms. The architecture defines what a credential validator must do (accept a submitted credential and return a pass/fail decision) without prescribing how enrollment is managed.

No single party possesses both a device identifier and a content hash for the same submission; the formal privacy properties in Section 4 formalize this guarantee for the base architecture.

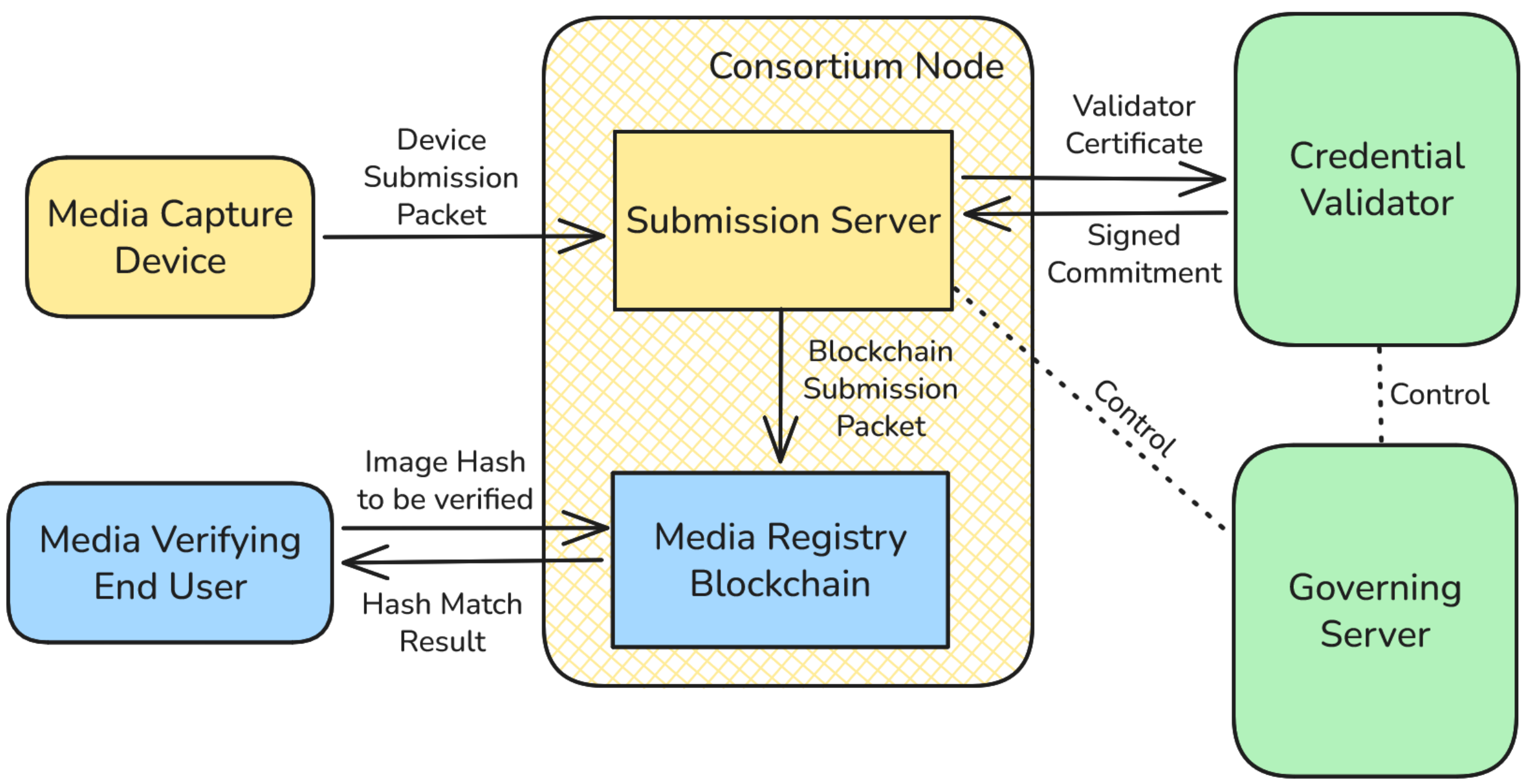

*Figure 1. System architecture overview. The Consortium Node (yellow) houses the Submission Server and Blockchain Registry. The Submission Server routes the Validator Certificate (credential channel) to the Credential Validator and the Blockchain Submission Packet (content channel) to the registry. The Credential Validator returns a Signed Commitment over the Media Record Hash. The Governing Server (green) issues control signals to both the Submission Server and Credential Validator. The Media Verifying End User queries the public registry directly.*

## 3.2 Data Channels

The architecture separates each submission into two independent data channels.

**The content channel** carries: a hash of the media content (H_content), derived by the capture device at the time of capture using any hash function that produces a stable output from the raw media data; an optional modification level indicator describing post-capture processing; and optional hashed metadata (timestamp, location) that enables selective disclosure without revealing values on-chain.

**The credential channel** carries: an encrypted device token (the device's hardware credential, encrypted under key material held exclusively by the credential validator); and a key reference identifying which decryption key table and index the validator should use. The base architecture uses ECIES: during device provisioning, the credential validator generates a key pair and loads the public key onto the device. At capture time, the device encrypts its credential under the validator's public key. Only the validator holds the private key and can decrypt. The device transmits both channels to the submission server in a single packet. The submission server separates them: it routes the encrypted credential to the validator and retains the content hash. The validator decrypts the credential, verifies device membership, and returns a signed approval. The validator never receives or observes the content hash; the submission server never decrypts the credential. This channel separation is the structural basis for the privacy properties proven in Section 4.

**Table 1.** Two-channel packet structure. Each channel carries only hash-substituted or encrypted data; neither channel individually constitutes a sufficient information set for device-to-content attribution.

| | Content Channel | Credential Channel |
|---|---|---|
| Carries | h(media); h(metadata) [optional] | Enc(device_id, k_v); key_reference |
| Does not carry | media content; device identifier | content hash; plaintext device_id |
| Routed to | Registry (public) | Credential Validator (private) |
| Sufficient for attribution | No | No |

**Table 2.** Single-channel possession is insufficient for attribution. Each party's observable information is semantically non-assembled: no party holds a sufficient information set for the target identification function.

| Party | Observes | Cannot determine |
|---|---|---|
| Registry observer | Content hashes; timestamps | Which device produced which content |
| Submission server | Both channels in transit | device_id (token encrypted under k_v) |
| Credential Validator | Enc(device_id, k_v) | Which content was authenticated |
| Adversary controlling one channel | One attribution surface | The other; correlation requires both simultaneously |

### 3.3 Registry Storage

The system never receives, stores, or processes media content at any stage. Once the validator's approval is received, the submission server posts to the registry: the content hash, the approval signature, and a posting timestamp rounded to a privacy-preserving interval. The registry record contains no device identifier, no manufacturer identifier, and no information that allows an observer to determine which device produced which content.

A consortium blockchain is used rather than a simpler append-only transparency log (such as Certificate Transparency) because append-only ledgers alone do not protect against compelled censorship. A single-operator transparency log can be compelled to remove or suppress specific entries by any locally enforceable legal order. A consortium blockchain distributes validation across multiple independent organizations: removing a record requires compelling a threshold of validators simultaneously, across organizational and jurisdictional boundaries. The strength of this property depends on the anti-concentration controls described in Section 7.2; a validator set with shared ownership or common legal jurisdiction would degrade it. For journalism and public-interest media contexts, where compelled removal of authentication records is a credible operational threat, this censorship resistance property is a design requirement rather than an incidental feature.

### 3.4 Verification

Public verification requires only the media content. A verifier computes the content hash locally and queries the registry directly. The registry returns a record if one exists. No subscription, credential, or trusted intermediary is required. The verifier learns whether the content was authenticated by a registered device; they learn nothing about which device.

The registry is non-self-describing and semantically inert to observers who do not already possess candidate media. The registry intentionally does not support attribution search, reverse lookup, or photographer enumeration. Querying the registry yields content hashes with no attributable metadata.

# 4 THREAT MODEL AND FORMAL PRIVACY PROPERTIES

Semantic Non-Assembly is enforced through intentional context blindness: each component is structurally limited to observing only the attribution-surface fragment required for its function, making the sufficient information set unassemblable below the compromise threshold. Privacy holds through structural non-assembly, not through operator trust. Channel separation structurally prevents any single component from holding both a device identifier and a content hash; where protocol-level separation is insufficient (principally, the prohibition on maintaining device-to-token mapping records outside the system), compliance is enforced through publicly auditable consortium governance rather than through the protocol itself. Property A (Device Non-Correlation) is established by an executable ProVerif proof. Properties B and C (Registry Observer Non-Identification and Submission Server Blindness) are verified by observational equivalence in BirthmarkBC.pv, using the same biprocess technique as Property A. Active defense gate properties (Property D) are formally verified in BirthmarkD.pv and documented in Appendix A.4. All properties are scoped in Appendix A.

## 4.1 Adversary Model

**Network Adversary (A_NET).** Controls network infrastructure between system components, with the ability to intercept, modify, replay, and inject messages on public channels. Submission servers may be compromised: a compromised server can inspect routed data but cannot decrypt encrypted credential tokens (only the credential validator holds decryption keys) and cannot forge validator signatures. Up to 33% of registry validator nodes may be compromised; honest validators retain the majority required for record finalization above this threshold. Simultaneous compromise of multiple independent components requires coordinated compromise across organizational boundaries and is outside the assumed adversary model.

Liveness under sub-threshold non-participation is handled by the consortium governance layer through validator removal, not by the consensus protocol alone. Governance removal is a prospective cleanup action, not a recovery mechanism. During the interval between a validator becoming non-participating and governance removal completing, submissions continue to finalize as long as the remaining honest validators exceed the 2/3 threshold; the non-participating validator's absence reduces the safety margin but does not interrupt service. Submissions requiring dual-approval from a non-participating server are queued or rerouted to an alternate submission server per deployment policy.

**Hardware Adversary (A_HW).** Has physical access to capture devices, with the ability to attempt credential extraction from hardware secure elements via invasive physical techniques. Such techniques are costly, time-intensive, and affect only individual devices, making bulk compromise economically infeasible under current secure element certification standards (EAL5+). Forging device credentials is also within scope. Replicating manufacturing-calibrated hardware fingerprints requires physical access to manufacturing equipment; no practical attack is publicly known under current manufacturing assumptions, though this is an empirical observation rather than a formal bound. Successful extraction affects only the compromised device and any key table entries it shares with other devices.

**Privacy Adversary (A_PRIV).** Seeks to correlate specific capture devices to specific verified content records. Capabilities include: observation of all public registry records; monitoring of network traffic including timing and packet sizes; access to any single system component (validator, submission server, or registry) but not multiple components simultaneously. Cannot break AES-256 encryption or SHA-256 collision resistance. Legal compulsion or coercion of individual operators is an out-of-scope threat. Single-operator compulsion is addressed through mandatory governance removal rather than protocol-level resistance; simultaneous multi-jurisdictional compulsion requires the same coordinated multi-component access as other out-of-scope attacks. The architecture requires that credential validators and submission servers operate under continuously audited structural constraints whose satisfaction is a deployment requirement enforced through governance rather than through the protocol itself. The constraints are publicly auditable through the consortium governance layer: validator node operators are publicly registered, enabling inspection of the validator set for shared ownership, board membership, or common legal jurisdiction. Constraint satisfaction is thereby an auditable deployment property rather than an invisible assumption.

**Out of scope.** Correlation across both channels requires concurrent access to the credential validator's decryption keys and the submission server's routing records; this adversary class is documented in Section 4.6.

### 4.2 Design Goals

**G1: Device-Rooted Authenticity.** The system must cryptographically prove that content originated from a registered capture device. Adversaries cannot generate valid verification records without possessing a legitimate registered device or successfully breaking hardware security mechanisms.

**G2: Privacy Preservation.** No single party can correlate a specific capture device to specific verified content. Privacy is computational: it holds against adversaries who observe any single system component but not against adversaries who simultaneously compromise both the credential validator and submission server.

**G2a: Extreme Data Minimization.** As a structural corollary of G2: each system component is designed to hold only the data necessary for its function, such that no individual component holds sufficient information to expose creator-to-content mappings. G2 is achieved structurally through the two-channel separation: no component in the base architecture persistently assembles the relationship between a specific device and specific content.

**G3: Metadata Independence.** Verification records must survive format conversion, compression, and platform reprocessing. Verification requires only the media content itself.

**G4: Public Verifiability.** Any party must be able to verify content authenticity without credentials or intermediary, without revealing the verifier's identity.

**G5: Minimal Storage.** Registry records are compact (153 bytes on-chain), enabling low-cost validator operation. Compactness also serves a privacy function: minimal on-chain footprint reduces the fingerprinting surface available to observers correlating submission timing, record size, or metadata patterns.

**G6: Temporal Data Minimization.** Forensic capability is time-limited. All transaction log records, rotation schedules, nonces, and window decryption keys are automatically deleted after three years from window close. After expiry, no forensic traversal is possible regardless of component compromise.

### 4.3 Device Non-Correlation

**Property:** The credential validator learns the device identity when confirming a credential, but receives no content-hash information for that submission. Formally: given device $D_i$ authenticating content items $C_1, C_2, \dots, C_n$, the validator observes authentication events but receives no information sufficient to correlate device identifiers with content hashes within the defined adversary model. Content hashes are never transmitted to the validator. The formal basis is established by the observational equivalence proof in Appendix A.2.

### 4.4 Registry Observer Non-Identification

**Property:** Registry observers see content hashes and coarsened posting timestamps but cannot identify devices, credential validators, or content creators. Given registry record R containing only content_hash and timestamp, observer O cannot determine which device created R with probability greater than random guessing over the registered device population. The formal basis is established by the observational equivalence proof in BirthmarkBC.pv (Appendix A.3).

### 4.5 Submission Server Blindness

**Property:** Submission servers receive the full submission packet but cannot identify specific devices. Given encrypted credential token T from device $D_i$, the submission server cannot determine device_id with non-negligible probability.

T = ECIES(device_id, pk_v) where pk_v is the credential validator's public key, provisioned onto the device at registration. The corresponding private key sk_v is held exclusively by the credential validator and never transmitted to the submission server. S cannot determine device_id with non-negligible probability by standard reduction to the IND-CCA2 game for ECIES. S receives only a binary approval decision from the validator and never receives the decrypted token value. BirthmarkBC.pv verifies this property directly; the ProVerif model's aenc/adec primitives map to ECIES in the reference instantiation without gap.

### 4.6 Composite Guarantee

The three properties together mean that correlation of a specific device to specific content requires simultaneously compromising the credential validator's decryption keys and the submission server's routing data. The base architecture achieves semantic non-assembly at t=1: any single-component compromise yields only attribution-surface fragments. The back-lookup extension makes full sequential traversal of all four forensic chain parties the only attribution path; no partial coalition of fewer than four parties can shortcut the chain. Independence here means the credential validator and submission server must not share ownership, legal jurisdiction, or key material; the structural controls enforcing this assumption are documented in Section 7.2. Under standard blockchain fault tolerance assumptions (requiring honest validators to outnumber compromised ones by at least 2:1), this attack surface is manageable for practical threat models.

The privacy guarantee is computational rather than information-theoretic. An adversary who compromises both the validator and submission server infrastructure can perform correlation.

## 5 THE REVOCATION TRADEOFF

### 5.1 Back-Lookup: A Four-Party Forensic Extension

For contexts where device revocation capability is required, the architecture supports a back-lookup extension structured as a four-party sequential forensic path. Each step in the path requires the previous one; there is no shortcutting. SS1 and SS2 are positional role labels (any submission server may occupy either) indicating which received the original submission (SS1) and which holds the relevant window decryption keys (SS2). The four parties and their roles are:

**Governing Server (GS).** Holds per-rotation-window nonces. The submission server encrypts its key rotation log with a nonce, transmits the nonce to the GS, and deletes the nonce locally. The GS is the only party that can make the key rotation log intelligible. Access to the GS is required to begin any forensic traversal.

**Submission Server 1 (SS1).** Holds the key rotation log in encrypted form (unintelligible without the GS nonce). The log records which submission server holds the decryption key for each rotation window's transaction records. With the GS nonce, SS1 can identify which SS2 to contact for a given transaction.

**Submission Server 2 (SS2).** Holds decryption keys for transaction logs of the rotation windows it was assigned. With a request from SS1 (which required the GS nonce to identify SS2), it decrypts the relevant transaction log to identify the specific transaction.

**Credential Validator (CV).** Holds device credentials and matches them to transaction IDs. The CV is never queried directly in legitimate forensic operation. Instead, identified TxIDs are posted to the public revocation list; the CV reads this list autonomously and acts internally to blacklist the device.

Once a TxID is identified, it is posted to the public revocation list. The CV acts on this list internally, resolving device identity, revoking future submissions from that device, and initiating the window-level broadcast purge.

The privacy cost of this extension is bounded: the four-party chain must be traversed in full to enable attribution of any specific transaction. No single-component or two-component compromise enables attribution. The base architecture's non-attribution property does not hold for deployments that include back-lookup, but the multi-party sequential dependency ensures attribution requires coordinated access across four independent organizational boundaries.

### 5.2 Active Defense Mechanics

The four-party forensic path, when deployed, introduces an active defense layer with two kill switches operating at different points in the chain and a key rotation mechanism that bounds retrospective exposure. These mechanics apply only to deployments using the back-lookup extension; the base architecture does not require them. The kill switches are asymmetric in scope and reversibility; the correct mechanism depends on the nature of the threat.

**Sliding-window key rotation and key provisioning.** Each submission server holds a physically provisioned master key, loaded directly onto hardware and never transmitted over a network. At the start of each rotation window, the server derives a window-specific key pair from its master key using a per-window random salt distributed by the GS. The derived encryption key is sent to whichever SS has been assigned to hold the decryption capability for that window. The GS generates and distributes the salt alongside the window assignment schedule but does not retain the salt, ensuring the GS accumulates no key material. The window assignment schedule is randomized each rotation to prevent pairing predictability. Window decryption keys are always distributed to submission servers operating under a different legal jurisdiction than the originating server; no submission server holds decryption keys for windows it originated, and no two servers in the key relationship share a jurisdiction.

Key tables are valid for a defined rotation window, nominally 30 days. Each submission server maintains an encrypted key rotation log recording which server holds decryption keys for each window. Upon rotation, the nonce for the retiring window is destroyed at the GS; the originating server's encrypted log entry for that window becomes permanently unintelligible. Key rotation does not affect the registry: content hashes written to the blockchain are permanent and unaffected by rotation.

**Kill switch: Emergency Blinding Protocol (EBP).** Triggered by the Governing Server via the Emergency Purge Broadcast, following confirmed forensic traversal or automated Decryption Key Tripwire detection. When a device has been identified via the four-party chain, its TxID is posted to the public revocation list. The CV acts on this list internally, revokes the device, and executes the mandatory purge instruction broadcast by the GS. The purge destroys all TxID to Device_ID associations for the entire compromised rotation window across the system. This is a window-level operation: all forensic paths for all transactions in that window are permanently severed, not just the confirmed fraudulent transaction. The window purge is irreversible; it trades the forensic traceability of an entire window for permanent protection of all attribution paths within it. The constraint preservation model governing the prohibition on device-to-token mapping records is described in Section 7.2; a concrete hardening example for that constraint is also provided there. The GS is the tripwire's failure point: an adversary controlling the GS can suppress the purge notification before it fires, making GS integrity the highest-value assumption in the active defense chain.

Targeted Privacy Attack
Stage 1: Submission Server 1 (SS1)
Attacker knows target image hash and approximate submission time window
Attacker gains SS01 endpoint access
Nonce intercepted at endpoint
Reads encrypted key rotation log
Automatic: All key rotation log queries are reported to the Governing Server
Use nonce to decrypt key rotation log to identify location of decryption key
Stage 2: Submission Server 2 (SS2)
Attacker gains SS08 endpoint access
Retrieves decryption key from key storage
Automatic: All decryption key queries are reported to the Governing Server
Automatic message sent to Governing Server
Governing Server: Emergency Blinding Protocol
Governing Server receives report and checks report against planned forensic lookups. If no match is found:
Key Rotation Log Report - Send alert to on-call personnel
Decryption Key Storage - Triggers Emergency Blinding Protocol that signals all CVs to delete all transaction records within the indicated time window.
Attacker tries to find transaction record that contains the device token
Stage 3: Credential Validator
Attacker gains CV endpoint access
Transaction record for target image hash permanently deleted
No correlating data possible

*Figure 2. Targeted privacy attack traversal and automated interception. An attacker with sequential access to SS1, SS2, and CV must traverse the four-party chain in order. Key rotation log queries (Stage 1) and decryption key queries (Stage 2) are automatically reported to the Governing Server. When Stage 2 completes without a matching authorized*

*forensic lookup, the GS triggers the Emergency Blinding Protocol before the attacker can reach Stage 3. The transaction record is permanently deleted before CV endpoint access yields any correlating data.*

**Temporal expiry.** All components in the forensic chain automatically delete their associated data three years after window close: each submission server purges its encrypted transaction log entries and any decryption keys held for expired windows, the GS destroys associated nonces and schedule records, and CVs purge transaction records for expired windows. Deletion is distributed and independent (each component acts on its own schedule without coordination). Because each stage holds only partial traversal information, deletion of any single component's window data irreversibly severs the end-to-end attribution chain. The public registry is unaffected: content hashes and transaction IDs written to the blockchain are permanent records of authenticity and serve verification, not attribution.

# 6 REFERENCE INSTANTIATION: THE BIRTHMARK STANDARD

This section demonstrates that the architecture is deployable on constrained hardware under real operational constraints; it is not a specification of the Birthmark Standard, which is documented in full in [4]. The Birthmark Standard instantiates the two-channel provenance architecture for photographic media in journalism contexts, applying Semantic Non-Assembly to a deployment where ZK-based approaches are computationally infeasible and metadata-dependent systems fail at distribution. The reference instantiation includes the back-lookup forensic extension described in Section 5, with the privacy cost characterized in Table 3.

Deployment viability on constrained hardware is established by the architecture's computational requirements: device-side operations require ECIES key encapsulation (one elliptic curve scalar multiplication, approximately 1–5ms on ARM Cortex-A class processors), AES-256-GCM, and SHA-256, all available on virtually all modern embedded processors, with total device-side overhead well within a 100ms capture-path budget [11][12]. Registry records are 153 bytes on-chain, enabling nonprofit organizations to operate validator nodes at low cost. The architecture's governance instantiation is provided by journalism organizations, fact-checking networks, and press freedom advocates operating under the anti-concentration controls described in Section 7.2. Full implementation details, hardware benchmarks, on-chain record structure, and governance specification are documented in [4].

# 7 DISCUSSION

## 7.1 Relationship to Cryptographic Blinding Approaches

ZK-proof-based device authentication provides stronger protocol-layer guarantees: Device Non-Correlation holds even against a compromised credential validator. For deployment contexts where ZK infrastructure is available and capture-path latency permits, ZK approaches are preferable on that dimension. ZK systems carry computational hardness assumptions, trusted-setup requirements, and proof-generation implementation complexity on constrained hardware. This architecture substitutes coordinated multi-component organizational compromise as the threshold for failure, a categorically harder attack to execute under the dominant real-world breach modes. Real ZK deployments also generate operational metadata outside the formal model: network traffic patterns, validator logs, and timing correlations that are structurally indistinguishable from those of this architecture. Neither system eliminates operational metadata leakage; the difference lies in the consequences of leakage, not its occurrence. On constrained capture hardware, the performance gap is also a practical constraint: ZK proof generation via multi-scalar multiplication (MSM) on ARM-class embedded processors requires seconds to minutes per proof [5], introducing capture-path latency and thermal load incompatible with continuous-capture operation. Device-side operations in this architecture require ECIES key encapsulation, AES-256-GCM, and SHA-256, benchmarked within a 100ms total capture-path budget on commodity embedded hardware [11][12], with no thermal accumulation across successive captures. This architecture additionally provides a semantic minimization property ZK systems do not inherently provide: the system holds no media content, so no party can perform content-based attribution regardless of protocol-layer guarantees.

ZK-based device authentication and SNA are orthogonal. ZK addresses whether a device can prove membership without revealing which member it is; SNA addresses whether any component can assemble the sufficient information set for the target identification function. A ZK-based system does not automatically satisfy SNA, and this architecture satisfies SNA without ZK. For deployment contexts where both are achievable, they are complementary.

## 7.2 Bounded Disclosure Under Compromise

The separation of privilege principle applies directly [7]. Table 3 formalizes bounded disclosure under compromise. All rows in the table are subject to the three-year temporal expiry described in Section 5.2: after that point, no forensic traversal is possible regardless of compromise level.

**Table 3.** Bounded disclosure under compromise for the four-party forensic extension (Section 5.1). The base architecture rows (registry, individual servers, CV) apply to all deployments. The four-party chain rows apply only to deployments using the back-lookup extension. The system never holds media content; no compromise scenario enables passive bulk attribution without the adversary independently possessing candidate media. Legitimate forensic operation never queries the CV directly; any direct CV attribution query is definitionally adversarial and triggers the CV read tripwire, which notifies the GS and initiates the automated window purge.

| Component | Adversary gains | Cannot determine | Surveillance class |
|---|---|---|---|
| Registry (public) | Content hashes; coarsened timestamps | Device identity; photographer identity; any attribution | None |
| GS alone | Per-window nonces | Encrypted key rotation log; any transaction or device data | None |
| SS1 alone | Encoded key rotation log (unintelligible without GS nonce) | Which SS2 to contact; any transaction or device data | None |
| SS2 alone | Decryption keys for assigned rotation windows | Which transactions to decrypt (requires SS1+GS) | None |
| CV alone | Device credentials; membership database | Content hashes; which content any device authenticated | Trial-and-error clustering only; no content attribution |
| GS + SS1 | Decoded key rotation log; identity of SS2 per window | Transaction records (requires SS2); device identity (requires CV) | Identifies SS2; no attribution data |
| GS + SS1 + SS2 | Specific transactions identified within rotation window | Device identity (requires CV); CV query triggers tripwire | Transaction identification only |
| All four (adversarial) | Attempted attribution via direct CV read; CV tripwire fires before match | Passive bulk attribution (system holds no media) | Worst-case: targeted hypothesis testing on possessed media only |
| All four (legitimate) | TxID identified; device revoked; window purge executed | N/A | Device revoked; window purge severs all forensic paths for that window |

A compelled single infrastructure component yields only the bounded disclosure characterized in Table 3. Coalition Member compulsion triggers mandatory removal instead of retaining a compromised validator (per the reference instantiation's governance model), operationalizing graceful degradation under coercion: localized legal compulsion cannot silently become network-wide subversion, and the 67% supermajority required for finalization means a compelled minority cannot suppress records while the removal process completes. Operators in different jurisdictions are subject to different legal orders; simultaneous multi-jurisdictional compulsion is categorically harder to execute than single-operator compulsion and represents the same coordinated multi-component class of attack described as out of scope in Section 4.1.

Legal compulsion that requires a validator or submission server to violate the architecture's structural constraints constitutes a governance compliance failure. The compelled operator faces a binary choice: resist the order and remain in the consortium, or comply and be removed from the accepted server list. Either outcome is publicly observable through the audit record, converting geopolitical legal pressure from a silent compromise vector into a visible governance event.

The audited constraint preservation model treats this explicitly: the consortium governance layer provides continuous public auditability of constraint satisfaction, so any degradation of structural independence is potentially detectable through the public audit record before it becomes exploitable, though the audit mechanism provides a detection opportunity, not a detection guarantee, and an adversary with sufficient control over the governance process could suppress the audit signal. The governance layer does not enforce honesty directly; it continuously exposes violations of the structural constraints on which Semantic Non-Assembly depends. The three-year automatic expiry of all forensic chain data provides a hard temporal bound: regardless of what an adversary accumulates before expiry, the data required to complete a forensic traversal is permanently unavailable after that point. These measures reduce the passive observation value of the registry and raise the coordination cost of correlation attacks. They do not constitute cryptographic guarantees. Timing analysis, traffic analysis, validator behavior leakage, and operational metadata correlation remain partially possible.

The reference governance model incorporates explicit anti-concentration controls intended to preserve the structural independence assumptions on which Semantic Non-Assembly depends. Validator ownership is capped per organization (Section 3.3 of the governance charter), with related entities treated as a single governance actor for ownership-limit purposes. This converts distributed governance from an aspiration into an enforceable structural property, treating governance capture as a first-class adversarial concern. Excluding organizations with direct media-production or commercial content interests from validator governance (Section 3.1) reduces conflict-of-interest risk to censorship-resistance and provenance claims; credential validators (camera manufacturers) operate outside Coalition governance by design, accepting that manufacturer independence from editorial governance is preserved at the cost of no formal manufacturer voice in protocol decisions. These measures do not cryptographically guarantee long-term institutional independence, but they increase the operational difficulty of silent consolidation and reduce incentives for attribution-capability convergence. The governance layer is therefore a load-bearing component of the security architecture, not an administrative overlay.

One concrete hardening mechanism for this constraint is worth documenting here. Before connecting to the consortium network, a credential validator operator's manufacturing database is audited by an independent party who directly inspects the schema to confirm it supports only membership queries, with no device-identifier-to-token mapping present. The auditor and manufacturer co-sign the schema attestation, which is posted to the consortium blockchain. The attestation is immutable; any later audit triggered by suspicion can compare the current schema against the committed record. Discrepancy is provably post-certification, establishing clear accountability without requiring recurring audit structure. This mechanism is a hardening example for the audited constraint preservation model; the architecture does not prescribe it as the only compliant approach.

### 7.3 Limitations

**Scope: hardware provenance only.** Authentication records confirm that content originated from a registered device. They do not verify scene truthfulness, chain of custody after capture, or editorial integrity. Staged content from a registered unmodified camera passes verification. This scope is intentional.

**Worst-case failure mode.** In deployments using the back-lookup extension, coordinated compromise of all four parties in the forensic chain enables targeted hypothesis testing on candidate media already possessed by the adversary. The automated CV tripwire (Section 5.2) is the primary technical defense: it triggers the window purge before the adversary can complete the final step. If the tripwire fails or is disabled, the adversary can confirm whether a specific device authenticated specific content they already possess. This is not passive bulk attribution; the system holds no media, so mass deanonymization is not possible regardless of compromise level.

**Governance capture.** The anti-concentration controls in Section 7.2 are designed with governance capture as a first-class concern, as both an adversarial attack and an institutionally realistic failure mode: journalism-adjacent governance structures have historically been subject to capture by state-adjacent or commercially-aligned actors. Validator ownership caps, jurisdictional distribution requirements, and the exclusion of commercial content interests from governance are direct mitigations. They reduce the incentive and opportunity for capture but do not eliminate it; the audited constraint preservation model is the detection mechanism if capture occurs.

**Formal model scope.** The ProVerif proofs in Appendix A establish channel separation properties under a network adversary model; the full scope of what the proofs cover and exclude is documented in Appendix A.5.

**Back-lookup retention risk.** In deployments that include back-lookup, the submission server's transaction ID mapping is a residual correlation surface. Compromise of this mapping enables an adversary to flag legitimate transactions as fraudulent or to monitor revocation patterns. Device identity is not stored on the submission server and requires additional access to the credential validator's internal records. The public revocation list records only opaque transaction identifiers; without SS1 and GS access to decode the key rotation log, a revocation entry cannot be associated with specific content, a device, or a submission event.

**Adoption dependency.** The system provides positive verification only. Content from devices whose manufacturers have not registered as credential validators lacks registry records and passes through the system unverified. Coverage is bounded by manufacturer participation, not by the architecture itself. In regulatory or legal contexts, this creates a structural asymmetry: content authenticated through participating manufacturers carries positive verification while content from non-participating hardware does not. This two-tier dynamic is not created by the architecture (unverified content exists in any system that provides positive verification only), but the architecture does not eliminate it either. Organizations using non-participating hardware are in no worse a position than they would be without the system, but they do not benefit from it. The natural mitigation is open-source credential infrastructure that allows smaller manufacturers and device categories to participate at low cost;

the architecture does not prescribe this but is compatible with it. Deployers operating in regulatory contexts where verification asymmetry could have legal consequences should assess participation coverage before deployment.

**Content hash collision.** Verification records are bound to content hashes, not content. Two distinct media files with the same SHA-256 hash would share a verification record. At realistic submission volumes (10^9 per day), accidental collision probability remains negligible under SHA-256's 2^128 collision resistance. Constructing a perceptually coherent image that intentionally collides with a target hash is computationally infeasible under current assumptions; this attack class is outside the defined adversary model.

**Hardware assumptions.** The security of device credentials depends on hardware secure element integrity. Physical extraction of credentials from EAL5+-certified secure elements is costly and affects only individual devices; bulk compromise is economically infeasible under current certification standards. This is an empirical observation rather than a formal bound and may change as hardware attack techniques evolve.

**Data classification under GDPR.** The architecture does not process personal data in the conventional sense: the registry holds only content hashes and coarsened timestamps, and no component retains device identifiers or photographer identity. In photojournalism deployment contexts, however, a content hash of a photograph of an identifiable individual (combined with a coarsened timestamp) may constitute personal data under GDPR Article 4(1) when held by a party that independently possesses the underlying media, since the hash is linkable to an identifiable person through that media. The architecture does not create this linkage, and no component holds both the hash and the media. Whether content hashes in isolation constitute personal data under the relevant jurisdiction's interpretation of GDPR is a deployment-specific legal question outside the scope of this paper; deployers operating under GDPR should obtain appropriate legal assessment for their specific context.

## 8 CONCLUSION

We have introduced Semantic Non-Assembly: a class of privacy guarantee characterized by the information yield of component exposure rather than the difficulty of achieving it. Restricting access to an existing store is a weaker property: it depends on the store existing and the restriction holding. In the base protocol, the guarantee is structural rather than procedural: architecture enforces it; policy cannot. Deployments using the back-lookup extension combine this structural guarantee with audited organizational constraints, as documented in Section 5. The guarantee is distinct from confidentiality, access control, information flow control, and statistical disclosure limitation. It applies to systems where a target identification function is authorized but must not be passively available. A two-channel provenance architecture instantiates the guarantee with ProVerif-verified formal properties. The Birthmark Standard demonstrates deployability on constrained capture hardware.

Structurally, Semantic Non-Assembly holds: no component ever holds sufficient information to correlate a device to content, and the system never holds media content at any stage. In the base architecture, even fully colluding operators cannot perform passive bulk attribution or photographer enumeration without independently possessing the underlying media; deployments using the back-lookup extension support targeted attribution via the four-party forensic chain, with the privacy cost bounded as documented in Section 5. Under the dominant real-world breach modes, each component compromise yields only attribution-surface fragments; worst-case coordinated subversion produces targeted confirmation on already-possessed media, not generalized deanonymization.

Both tradeoffs in this architecture are explicit and documented. Audited constraint preservation does not protect against simultaneous compromise of the credential validator and submission server. Back-lookup capability, when deployed, partially breaks Device Non-Correlation for the credential validator. Neither limitation is hidden, and the precise scope of each is stated in Sections 5 and 7.3.

Deployment viability follows from the architecture's computational requirements: device-side operations require ECIES key encapsulation, AES-256-GCM, and SHA-256, within a 100ms capture-path budget on commodity embedded hardware. A reference instantiation (the Birthmark Standard) demonstrates 153-byte on-chain records and governance by journalism organizations operating on commodity hardware at low per-node cost. The specification and reference implementation are published under Apache 2.0.

In a threat environment where leaks are inevitable, designing for the consequences of failure is at least as important as designing against it.

**Repository:** https://github.com/Birthmark-Standard/Birthmark

## APPENDIX A: PROVERIF FORMAL VERIFICATION

### A.1 Overview

This appendix presents the formal verification of the privacy properties described in Sections 4 and 5.2 using ProVerif, a cryptographic protocol verifier that uses a process algebra (applied pi-calculus) to model protocol participants and prove security properties. The model and proof for Property A (Device Non-Correlation) are adapted from Ryan (2026) [4]. Properties B and C are verified by observational equivalence in a second executable ProVerif model (BirthmarkBC.pv), documented in Appendix A.3. Formal verification of the active defense gate properties (BirthmarkD.pv) is documented in Appendix A.4.

The ProVerif model captures the core separation property: the Device sends the content hash and encrypted credential to the Server; the Server routes the encrypted credential to the Validator and posts the content hash to the registry; the Validator returns a binary decision without observing the content hash. The model omits optional implementation details such as timestamp rounding and key table indexing; neither affects the channel separation property being proved, as both operate after or within a single channel rather than crossing channel boundaries.

ProVerif operates under the Dolev-Yao adversary model: the adversary controls the network, can intercept and inject messages on public channels, but cannot break cryptographic primitives (AES-256, SHA-256). The model treats cryptographic operations as secure by assumption rather than analyzing their bit-level implementation, consistent with the Network Adversary (A_NET) defined in Section 4.1.

Table A.1 summarizes the four verified properties, the ProVerif model used for each, and the proof technique applied.

| Property | Model | Technique | Guarantee |
|---|---|---|---|
| A | BirthmarkA.pv | Observational equivalence | Device Non-Correlation |
| B | BirthmarkBC.pv | Observational equivalence | Registry Observer Non-Identification |
| C | BirthmarkBC.pv | Observational equivalence | Submission Server Blindness |
| D | BirthmarkD.pv | Correspondence + reachability | Active Defense Gate |

### A.2 Property A: Device Non-Correlation (Executable Proof)

We model the protocol as three concurrent processes: Device, Server, and Validator. The Device encrypts its credential token under the Validator's public key and sends both the encrypted token and the binding value to the Server. The Server forwards the token to the Validator and, on approval, posts the content hash to the registry. The Validator decrypts the token, checks membership, and returns a decision without observing the content hash.

The proof establishes that two scenarios are indistinguishable to any outside observer: one in which the Device authenticates content item C_1, and one in which it authenticates C_2. If the Validator cannot distinguish these scenarios, Device Non-Correlation holds.

**ProVerif Model — Property A:**

```
(* Executable ProVerif model verifying Property A *)
type key. type skey. type pkey.
fun pk(skey): pkey.
fun aenc(bitstring, pkey): bitstring.
reduc forall m:bitstring, sk:skey; adec(aenc(m, pk(sk)), sk) = m.
fun hash(bitstring): bitstring.
free pub: channel.
free validator_private: channel [private].
free registry: channel.
free v_ok: bitstring.  free v_no: bitstring.
free sk_v: skey [private].  letfun pk_v = pk(sk_v).
free T1: bitstring [private].  free T2: bitstring [private].
free contentA: bitstring [private].  free contentB: bitstring [private].
let device(token: bitstring, content: bitstring) =
  let h = hash(content) in
  let enc_token = aenc(token, pk_v) in
  out(pub, (h, enc_token)).
let server =
  in(pub, (h: bitstring, enc_token: bitstring));
  out(validator_private, enc_token);
  in(validator_private, decision: bitstring);
  if decision = v_ok then out(registry, h).
let validator =
  in(validator_private, enc_token: bitstring);
  let token = adec(enc_token, sk_v) in
  if (token = T1 || token = T2) then out(validator_private, v_ok)
  else out(validator_private, v_no).
process out(pub, pk_v);
  ( (!device(T1, choice[contentA, contentB])) | (!server) | (!validator) )
```

**ProVerif Output:**

```
RESULT Observational equivalence is true.
--------------------------------------------------------------
Verification summary: Observational equivalence is true.
--------------------------------------------------------------
```

This result confirms that an adversary with full network observation capabilities cannot distinguish between scenarios where the same device authenticates different content items. The Validator receives only the encrypted credential token; the registry receives only the content hash. No channel carries both. Device Non-Correlation holds. Termination warnings in the full ProVerif output indicate selection heuristics used by the resolution engine and do not affect the validity of the result.

### A.3 Properties B and C: Registry Observer Non-Identification and Submission Server Blindness (Observational Equivalence Proof)

**ProVerif Model (BirthmarkBC.pv):**

```
(* BirthmarkBC.pv: Observational Equivalence Proof            *)
type skey. type pkey.
fun pk(skey): pkey.  fun aenc(bitstring, pkey): bitstring.
reduc forall m:bitstring, sk:skey; adec(aenc(m,pk(sk)),sk) = m.
fun hash(bitstring): bitstring.
free pub: channel.  free mfr_private: channel [private].
free blockchain: channel.
free v_ok: bitstring.  free v_no: bitstring.
free sk_m: skey [private].  letfun pk_m = pk(sk_m).
free T1: bitstring [private].  free T2: bitstring [private].
free img: bitstring [private].
(* Property B: registry view is hash(img) in both worlds.     *)
(* Observational equivalence checked via choice[T1,T2] below. *)
```

```
(* Property C: aenc(T1,pk_m) vs aenc(T2,pk_m): indistinguish- *) (* able under IND-
CCA2. Server cannot identify device.        *)
let camera(token: bitstring, image: bitstring) =
    let h = hash(image) in
    let enc_token = aenc(token, pk_m) in
    out(pub, (h, enc_token)).
let server =
    in(pub, (h: bitstring, enc_token: bitstring));
    out(mfr_private, enc_token);
    in(mfr_private, decision: bitstring);
    if decision = v_ok then out(blockchain, h).
let manufacturer =
    in(mfr_private, enc_token: bitstring);
    let token = adec(enc_token, sk_m) in
    if (token = T1 || token = T2) then out(mfr_private, v_ok)
    else out(mfr_private, v_no).
process out(pub, pk_m);
    ((!camera(choice[T1,T2], img)) | (!server) | (!manufacturer))
```

**ProVerif Output:**

```
--------------------------------------------------------------
Verification summary:
Observational equivalence is true.

--------------------------------------------------------------
```

The observational equivalence result establishes that no Dolev-Yao adversary can distinguish a world where device T1 authenticated img from a world where device T2 did. The two worlds differ only in which token is encrypted: the registry receives hash(img) in both, and the submission server receives aenc(T1, pk_m) vs aenc(T2, pk_m), which are indistinguishable under the symbolic model's treatment of asymmetric encryption. Since the Dolev-Yao attacker subsumes both the registry observer (Property B) and the submission server (Property C), the equivalence result establishes both properties simultaneously and at a stronger level than reachability: not only can the attacker not derive the tokens, it cannot distinguish which token was used at all.

### A.4 Property D: Active Defense Gate (BirthmarkD.pv)

BirthmarkD.pv verifies five properties of the active defense gate and four-party back-lookup chain under two process variants: GS in purge state (main process) and GS in normal state (sanity-check process). The main process grants the adversary two explicit capabilities: observation of the purge command, and possession of the opaque CV table record (senc(ss2_addr, record_nonce) and senc(key_index, record_nonce)). The sanity-check process adds gs_nonce_server, enabling chain completion and verifying D.3 and D.4 non-vacuously.

ProVerif Model (BirthmarkD.pv):

```
(* Symmetric authenticated encryption — models AES-256-GCM *)
fun senc(bitstring, bitstring): bitstring.
fun sdec(bitstring, bitstring): bitstring.
equation forall m: bitstring, k: bitstring;
  sdec(senc(m, k), k) = m.

free pub:          channel.
free cv_gs_chan:   channel [private].
free ss1_gs_chan:  channel [private].
free ss1_ss2_chan: channel [private].
free revoc_list:   channel [private].

free ok_token:     bitstring [private].
free record_nonce: bitstring [private].
free ss2_addr:     bitstring [private].
free key_index:    bitstring [private].
free window_key:   bitstring [private].
free table_record: bitstring [private].
free device_id:    bitstring [private].
free ss1_auth:     bitstring [private].
free purge_cmd:    bitstring.
free liveness_req: bitstring.

event GS_SentOK().          event GS_SentPurge().
event SS1_ReceivedNonce().  event SS1_DerivedIndex().
event SS2_ReleasedKey().    event BackLookupComplete().
event CV_AccessedTable().  event CV_ExecutedPurge().
event CV_RevokedDevice(bitstring).
```

```
(* D.1 *)  query event(CV_AccessedTable()) ==> event(GS_SentOK()).
(* D.2 *)  query attacker(table_record).  query attacker(device_id).
(* D.3 *)  query event(SS2_ReleasedKey()) ==> event(SS1_ReceivedNonce()).
(* D.4 *)  query event(BackLookupComplete()) ==> event(SS2_ReleasedKey()).
(* D.5 *)  query attacker(ss2_addr).  query attacker(key_index).
(* Non-vacuity *)
query event(CV_ExecutedPurge()).
query event(BackLookupComplete()).
query event(SS1_DerivedIndex()).

process
  out(pub, purge_cmd);
  out(pub, senc(ss2_addr,  record_nonce));
  out(pub, senc(key_index, record_nonce));
  ( (!gs_purge) | (!ss1) | (!ss2) | (!cv_lookup) | (!cv_revoke) )
```

ProVerif Output (main process, GS in purge state):

```
--------------------------------------------------------------
Verification summary:

Query event(CV_AccessedTable) ==> event(GS_SentOK) is true.

Query not attacker(table_record[]) is true.

Query not attacker(device_id[]) is true.

Query event(SS2_ReleasedKey) ==> event(SS1_ReceivedNonce) is true.

Query event(BackLookupComplete) ==> event(SS2_ReleasedKey) is true.

Query not attacker(ss2_addr[]) is true.

Query not attacker(key_index[]) is true.

Query not event(CV_ExecutedPurge) cannot be proved.

Query not event(BackLookupComplete) is true.

Query not event(SS1_DerivedIndex) is true.

--------------------------------------------------------------
```

D.1 (gate correspondence) is true: CV table access structurally requires prior GS authorization. D.2 (purge-state secrecy) is true for both table_record and device_id: neither is reachable when GS is in purge state. D.3 (chain dependency) and D.4 (chain integrity) are true vacuously in purge state; they are verified non-vacuously by the sanity-check process (GS in normal state). D.5 (opaque index secrecy) is true: an adversary holding the full CV table record cannot derive ss2_addr or key_index without record_nonce, which is held by GS and never appears on a public channel. The non-vacuity check not event(CV_ExecutedPurge) cannot be proved, confirming the purge branch is reachable by design. not event(BackLookupComplete) and not event(SS1_DerivedIndex) are both true, confirming the back-lookup chain is structurally severed in purge state.

### A.5 Scope, Limitations, and Computational Soundness

Properties B and C are verified by observational equivalence in BirthmarkBC.pv: no Dolev-Yao adversary can distinguish a world where device T1 authenticated the submitted content from a world where device T2 did. The Dolev-Yao attacker subsumes both the registry observer (Property B) and the submission server's observable inputs (Property C); the equivalence result establishes both properties simultaneously.

The symbolic model treats AES-256 and SHA-256 as perfect. Abadi and Rogaway [10] establish computational soundness transfer results for classes of symbolic models under suitable cryptographic assumptions: IND-CCA2-secure encryption and collision-resistant, preimage-resistant hash functions. The symbolic model used here falls within the scope of those results: ECIES provides IND-CCA2 security, and SHA-256 provides collision resistance and preimage resistance under standard assumptions. The symbolic proof transfers to a computational guarantee that no PPT adversary can distinguish device-content correlations with non-negligible advantage. The ProVerif model's aenc/adec primitives directly reflect the ECIES deployment; no gap exists between the verified model and the reference instantiation.

Outside the current formal guarantees: (1) cryptographic hardness by reduction; (2) physical realization variables under the symbolic model (side-channel attacks, power analysis, and timing leakage in the A_HW class) are treated as out-of-scope physical implementation concerns under standard symbolic cryptographic assumptions, not as properties of the protocol; (3) traffic analysis and timing correlation; (4) governance compromise and legal compulsion; (5) blockchain consensus and liveness; (6) key management correctness; (7) implementation bugs. Active defense gate properties (Property D: gate correspondence D.1, purge-state secrecy D.2, back-lookup chain dependency D.3, chain integrity D.4, and opaque index secrecy D.5) are formally verified in BirthmarkD.pv and documented in Appendix A.4. The gate proofs cover the back-lookup extension only; the base architecture does not require them. Temporal ordering guarantees are documented in [4].

### A.6 Future Verification Directions

Several extensions to the current formal model would strengthen the guarantees. A game-based proof would establish that the symbolic guarantees hold against probabilistic polynomial-time adversaries, by reduction to IND-CCA2 of the encryption scheme and preimage resistance of the hash function. This would replace the reliance on computational soundness transfer results with a direct reduction. A formal timing model could establish bounds on traffic analysis resistance under specific network assumptions, converting heuristic traffic analysis resistance claims into formal bounds.

The current proofs establish a formal foundation for the core channel separation property; they should be understood as a necessary but not sufficient component of a complete security argument. Properties B and C are established by observational equivalence in BirthmarkBC.pv, using the same biprocess technique as Property A. All three channel separation properties (A, B, C) are established by observational equivalence. The proof basis for Properties B and C is described in Sections 4.4 and 4.5 respectively.